\documentclass{article}
\usepackage{epsfig}

\begin{document}

\title{Algorithmic Complexity in Noise Induced Transport Systems}

\author{C.M. Arizmendi, J.R. Sanchez \\
Dpto de F\'{\i}sica, Facultad de Ingenier\'{\i}a, \\
Universidad nacional de Mar del Plata, \\
Av. J.B. Justo 4302,\\
7600, Mar del Plata,\\
Argentina}

\date{}

\maketitle

\begin{abstract} 

Time correlated fluctuations interacting with
a spatial asymmetry potential are sufficient conditions to give rise
to transport of Brownian particles.
The transfer of information coming from the nonequilibrium bath,
viewed as a source of negentropy, give rise to the correlated noise.
The algorithmic complexity of an object provides a means of
quantitating its information contents.
The Kolmogorov information entropy or algorithmic complexity
is investigated in order to quantitate the transfer of information that occurs
in computational models showing noise induced transport.
The complexity is measured in terms
of the average number of bits per time unit necessary to specify the sequence
generated by the system.

\end{abstract}

\newpage

Recently a lot of attention has been paid to simple stochastic models in
which some of the energy in a nonequilibrium bath is used to obtain a net
transfer of information. The best known examples of this
process are  {\it thermal ratchets} or {\it correlation ratchets} where a
nonzero net drift speed may be obtained from time correlated fluctuations
interacting with broken symmetry structures.  Recently the
study of a wide variety of ratchets has 
received much attention due to their general interest and possible
technological applications in modeling molecular
motors and other nanoscale and mesoscale systems. For a general review
of thermal ratchets the reader is referred to Ref.\cite{doering}.
 
As was demonstrated by M. Millonas~\cite{Millonas}, the net current is
obtained by means of a source or sink of negentropy (physical information)
that allows the engine to operate. Numerous existing studies have addressed the
question of the efficiency in ratchets~\cite{efi} but, despite this 
widespread interest none of them did it from the perspective of information
theory and, up to our knowledge, a quantitative  estimate of 
the amount of negentropy was never obtained. 
A possible reason that no effort has been done to study
the entropy transfer on thermal ratchets may be that they constitute far
from equilibrium, non-ergodic systems and usual entropy calculations rely
on probabilistic ensemble concepts that require energy minimization
procedures associated with equilibrium arguments. Algorithmic complexity
was a concept introduced by Kolmogorov to avoid the probabilistic
interpretation of Shannon that has been recently used to obtain the
information content of far from equilibrium systems such as proteins and
fractal growth processes \cite{dewey}. This characteristic makes
algorithmic complexity specially useful to obtain the variation of entropy
or physical information of far from equilibrium systems such as  thermal
ratchets. 

Following the idea developed by Crisanti et al \cite{crisanti} and Paladin
et al \cite{Paladin} that applied algorithmic complexity to characterize
the chaoticity of dynamical systems with noise, we study the algorithmic
complexity of thermal ratchet motion.   

The algorithmic complexity of an object is broadly defined as the length
in bits of the shortest description for that object \cite{dewey}. In other
words, complexity can be characterized as the length of the shortest
possible definition of the sequence itself.

In the framework of information theory, if one wants to transmit the
sequence, through a noisy channel, until it deviates from the
deterministic one a tolerance threshold $\Delta$, one can use the
following strategy:\par

Specify the initial condition $x(1)$ with precision $\delta_0$ using a
number of bits $n = log_2 (\Delta/\delta_0)$ which permits to arrive up to
a time $\tau_1$ where the error equals $\Delta$. Then specify again a new
initial condition $x(\tau_1+1)$ with a precision $\delta_0$ and arrive up
to a time $\tau_2$, and so on $N$ times. The number of bits necessary to
specify a sequence with a tolerance $\Delta$ up to $T_{max} = \sum_{i=1}^N
\tau_i$ is $\simeq N n$ and the mean information generated per time step
is $\simeq N n / T_{max} = K_\sigma/ln 2$ bits. Following the idea of
algorithmic complexity as the length in bits of the mean information,
$K_\sigma$ is the complexity associated to stochastic motion. Thus,
$K_\sigma$ can be obtained  as \cite{Paladin}:
\begin{equation}
K_\sigma = \frac{1}{\overline \tau} ln \left( \frac{\Delta}{\delta_0}
\right),
\end{equation}
where 
\begin{equation}
\overline \tau = \lim_{N \rightarrow \infty} \frac{1}{N} \sum_{i=1}^N
\tau_i.
\end{equation}

The model for ratchets that we have used is based on
the model proposed by Astumian and Bier
\cite{astu}. Two of the authors, recently used this model to study 
the approach to steady state in ratchets \cite{ariz} and to study 
ratchets with finite inertia\cite {ariz1}. 
It consists of an asymmetric piecewise linear potential where 
the barrier height fluctuates between two states. 
Astumian and Bier showed that the fluctuations of the barrier
height can produce a net flow even with the net
macroscopic force that is zero at all times.  
The general Langevin equation describing the ratchets
that we will consider is 
\begin{equation}
{dx \over dt} = - u'(x,t) + \sqrt{2k_B T} w(t),
\end{equation}
where $u'(x,t) =  {\partial u(x,t) \over \partial x}$, 
$T$ is the temperature and $k_B$ is the Boltzmann constant.
The potential $u(x,t)$ switches on and off the $v(x)$
asymmetric potential, and
$w(t)$ represents the Gaussian noise term with
delta function correlation
$ \langle w(t) w(t') \rangle = \delta (t-t') $. 

The fluctuations cause the potential $u(x,t)$ to
switch between 
$u_{+}(x) = v(x)$ and $u_{-}(x) = 0$ with a dichotomous Markov process
governed by the master equation
\begin{equation} {d \over dt} \pmatrix{P_+(t) \cr P_-(t) \cr} =
\gamma
\pmatrix{-1 & 1 \cr 1 & -1 \cr} \pmatrix{P_+ \cr P_-
\cr},
\end{equation} 
where $\gamma$ represents the frequency of the
flipping between $u_{+}(x)$ and $u_{-}(x)$.
 
In our numerical simulations we have followed the
approach of Elston and Doering \cite{elston}. Details of the simulation
procedure can be found in references \cite{ariz} and \cite{ariz1}.

In order to obtain the algorithmic complexity associated to ratchet
motion, two different walkers are allowed to move under the ratchet
potential with initial conditions separated by $\delta_0$, $x(1)$ and
$x'(1)=x(1)+\delta_0$. Both motions continue until the difference equals
$\Delta$ at time $\tau_1$. Then begin again with initial conditions
$x(\tau_1+1)$ and $x'(\tau_1+1)=x(\tau_1+1)+\delta_0$, and so on $N$
times.
	
Simulations were carried out on ratchets with an asymmetric 
barrier \cite{astu,ariz} switched on and off with different flipping rates.
The asymmetric parameter used was $\alpha = 10/11$ and the barrier height
was $E_0 = 10$. 
This is a very important kind of ratchet because the
experimental studies of ratchets done by
Rousselet et al \cite{rous} and by Faucheux et al \cite{mass} were carried
out with an asymmetric potential switched on and off
periodically and Ajdari and Prost~\cite{ajdari} used
the on-off barrier fluctuation as a method for
dielectrophosphoretic separation.

\begin{figure}[t]
\begin{center}
\psfig{figure=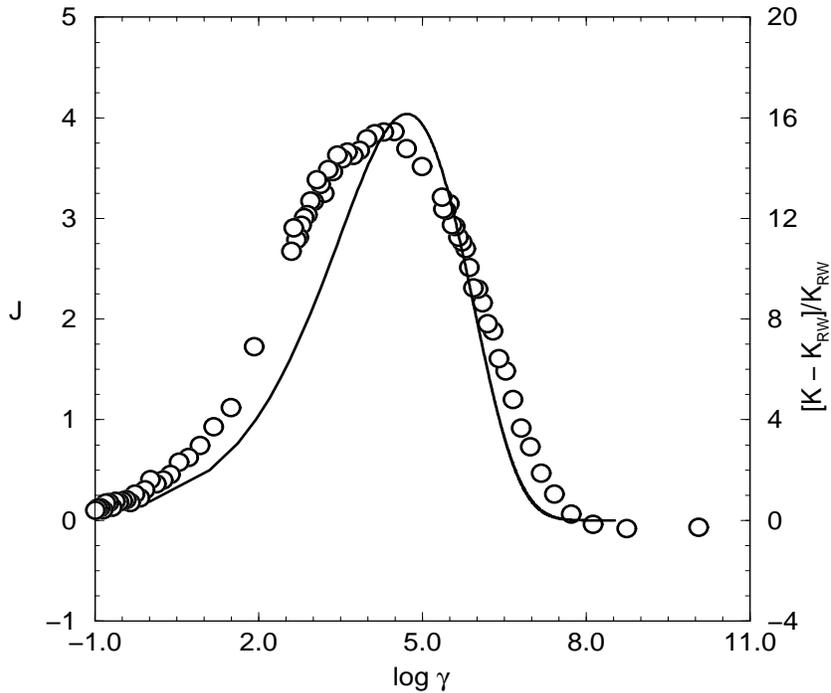,height=9cm,width=9cm}
\caption{\tt Comparison between absolute value of the current (solid line) and
relative complexity (circles) of an on-off ratchet vs logarithm of flipping rate.}
\end{center}
\end{figure}

In the limits of low and high flipping rates the current tends to zero and 
the system behaves essentially as a random walk. 
The relative complexity for a given flipping rate is defined as:
\begin{equation}
\frac{K - K_{RW}}{K_{RW}},
\end{equation}
where $K$ is the complexity at the given flipping rate and $K_{RW}$ is the
complexity for high flipping rate limit.
The relative complexity and the net current as functions of the logarithm of 
the flipping rate $\gamma$ are shown in Fig. 1.
In this figure it can be appreciated that the relative complexity
tends to zero in the limits of low and high flipping rates, where the current
tends to zero. On the other hand a dramatic increase of the relative
complexity appears  for flipping rates corresponding to the highest
currents.

The proportionality between $K$ and $J$ can be explained 
using the well known statistical mechanical relationship derived by
Zurek~\cite{zurek}
\begin{equation}
S = K + I ,
\end{equation}
\noindent where $S$ is the physical entropy, $K$ is the algorithmic complexity
and $I$ is the Shannon information entropy.
Zurek's relationship says that the physical or thermodynamic entropy of a system
is composed of two parts, that determined from the known information of the
system $K$ and that determined from the unknown or probabilistic information $I$.
When the microstates of a system are unknown $K = 0$ and the entire entropy
is due to the Shannon information, i. e., $S = I$. As observations are made on
the system the information content shifts from $I$ to $K$.

When there is a net current in the ratchet, the amount of known information
about the system increases with respect to the case of no net current. 
The existence of a net current implies there is some {\it more information 
known} about the system.
This amount of extra information comes from the negentropy supplied by 
the external nonequilibrium bath.
As a consequence, the maximum amount of extra information or complexity 
is obtained when the current is maximum.

Another way to characterize the behavior of complexity in on-off
ratchets is the scaling of the complexity with the error $\Delta$. We
studied the scaling behavior of complexity for the random walk limit
corresponding to a high flipping rate and for the highest current flipping
rate. The scaling behavior for both cases is shown in Fig. 2 where
$log(K)$ vs $log(\Delta)$ is plotted. In the random walk limit  $K \sim
\Delta^{-2}$ according to the standard diffusion law. For the highest
current flipping rate, due to the existence of a net current, 
the complexity scales as $K \sim \Delta^{-1}$.

\begin{figure}[t]
\begin{center}
\psfig{figure=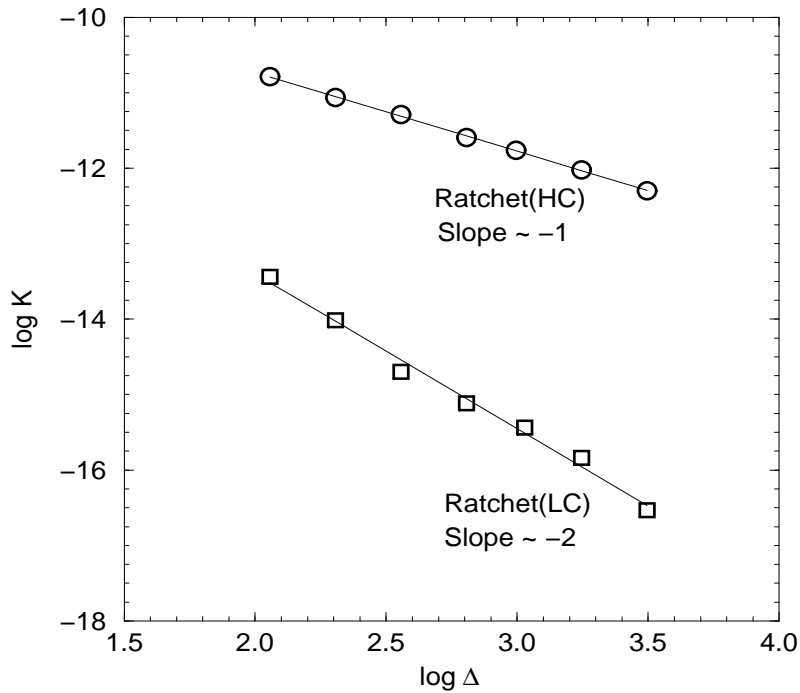,height=9cm,width=9cm}
\caption{\tt Log-log plot of the relative complexity vs the error
$\Delta$ for high and low current flipping rates.}
\end{center}
\end{figure}

The simulations were carried out using a rather high value of $\Delta
\simeq 10$ in order that the interaction between the ratchet potential and
the on-off fluctuation may take effect. Results were averaged 
over 500 independent runs. 

In conclusion, the net transfer of information from a source of negentropy 
that allows the net drift of particles in thermal ratchets may be obtained 
through the algorithmic complexity.
The negentropy transference results in an increase of complexity observed 
when high ratchet currents take place (see Fig. 1).
Additionally, a different scaling behavior of complexity at high currents
compared with the no current case, {\it i.e.} random walk limit, is
observed as can be seen in Fig. 2.
As far as we know, the algorithmic complexity lets us obtain, for the first
time, an estimate of the net transfer of information associated with ratchet motion.

\end{document}